\documentclass[showpacs,superscriptaddress,notitlepage]{revtex4-1}
\usepackage[english]{babel}
\usepackage{amsmath,amssymb}
\usepackage{amsfonts}
\usepackage{mathtools}
\usepackage{siunitx}
\usepackage{libertine}
\usepackage[usenames,dvipsnames]{color}
\usepackage{hyperref}
\usepackage{blindtext}
\hypersetup{colorlinks}
\hypersetup{linkcolor=black, citecolor=black, urlcolor=blue}
\usepackage[utf8]{inputenc}
\usepackage{color}
\usepackage{subcaption}
\usepackage[normalem]{ulem}
\usepackage{xcolor}
\usepackage{graphicx}

\newcommand{\dif}{\mathrm{d}}

\begin{document}

\title{Dynamic Stern layers in charge-regulating electrokinetic systems: three regimes from an analytical approach}

\author{B. L. Werkhoven}
\affiliation{Institute for Theoretical Physics, Center for Extreme Matter and Emergent Phenomena, Utrecht University, Princetonplein 5, 3584 CC, Utrecht, The Netherlands}
\author{S. Samin}
\affiliation{Institute for Theoretical Physics, Center for Extreme Matter and Emergent Phenomena, Utrecht University, Princetonplein 5, 3584 CC, Utrecht, The Netherlands}
\author{R. van Roij}
\affiliation{Institute for Theoretical Physics, Center for Extreme Matter and Emergent Phenomena, Utrecht University, Princetonplein 5, 3584 CC, Utrecht, The Netherlands}

\bibliographystyle{apsrev4-1}

\begin{abstract}
We present analytical solutions for the electrokinetics at a charged surface with both non-zero Stern-layer conductance and finite chemical reaction rates. We have recently studied the same system numerically [Werkhoven {\em et al.}, Phys. Rev. Lett. {\bf 120}, 264502 (2018)], and have shown that an applied pressure drop across the surface leads to a non-trivial, laterally heterogeneous surface charge distribution at steady state. In this work, we linearise the governing electrokinetic equations to find closed expressions for the surface charge profile and the generated streaming electric field. The main results of our calculations are the identification of three important length and time scales that govern the charge distribution, and consequently the classification of electrokinetic systems into three distinct regimes. The three governing time scales can be associated to (i) the chemical reaction, (ii) diffusion in the Stern layer, and (iii) conduction in the Stern layer, where the dominating (smallest) time scale characterises the regime. In the reaction-dominated regime we find a constant surface charge with an edge effect, and recover the Helmholtz-Smoluchowski equation. In the other two regimes, we find that the surface charge heterogeneity extends over the entire surface, either linearly (diffusion-dominated regime) or nonlinearly (conduction-dominated regime).
\end{abstract}

\maketitle

\section{Introduction}

While the field of electrokinetics is over a century old, interest in it has only grown since its foundations were laid by Helmholtz and Smoluchowski \cite{helmholtz,Smoluchowski}. In addition to applications in well-established fields such as geology \cite{geology} and catalysis \cite{catalysis}, the advent of micro- and nano-fluidics renewed interest in electrokinetics \cite{tagliazucchi,AjdariBoquet,nano1,nano2,nano3} due to applications in, for example, blue-energy harvesting \cite{blauw1,blauw2}. At the basis of all eletrokinetic systems is the interaction between fluid flow and a charge current. In a closed-circuit setup, an imposed voltage drop over a channel with a charged wall induces fluid flow via the electric body force in the Navier-Stokes equations, while in an open-circuit setup, an imposed pressure drop induces an electric field. This streaming electric field implies a voltage drop across the channel, the so-called streaming potential $\Delta \Phi$. In the linear response regime, the generated streaming potential is linearly related to the applied pressure drop $\Delta p$ via the Helmholtz-Smoluchowksi equation \cite{helmholtz,Smoluchowski},
\begin{equation}\label{eq:HSE}
\Delta \Phi=-\dfrac{\zeta \epsilon}{\eta G}\Delta p,
\end{equation}
where $G$ is the channel conductivity, $\epsilon$ and $\eta$ are the permittivity and shear viscosity of the liquid, respectively, and $\zeta$ is the zeta potential -- the electrostatic potential at the slipping plane of the charged surface of the channel. Since $G$, $\epsilon$ and $\eta$ are material properties of the liquid, measurements of $\Delta \Phi$ at a known $\Delta p$ allows one to use Eq. (\ref{eq:HSE}) to measure $\zeta$, an important surface property. The presence of the charged walls is essential for electrokinetic phenomena, since charged surfaces induce an Electric Double Layer (EDL) in the fluid adjacent to the surface. This diffuse layer of ions screens the charge of the surface. A fluid flow (electric field) through the charged EDL induces a charge current (body force), which in turn induces an electric field (fluid flow). A detailed understanding of the surface charge is therefore vital to describe electrokinetic systems accurately. 

The total channel conductivity is typically decomposed as $G=G_b+G_s/H$, with $G_b$ the bulk conductivity of the fluid, $G_s$ the surface conductivity, and $H$ the channel height \cite{lyklemaboek}. The significance of the surface conductivity is expressed by the Duhkin number Du=$G_s/(G_bH)$ \cite{lyklemaboek}. The surface conductivity $G_s$ can be further decomposed as $G_s=G_s^{EDL}+G_s^{S}$. Here, $G_s^{EDL}$ originates from the increased conductivity of the EDL with respect to bulk fluid, as first described by Bikerman \cite{bikerman1,bikerman2}, while $G_s^S$ is the conductivity due to the mobile charges in the Stern layer \cite{overbeekkruyt}, the quasi-2D layer in which the surface charge resides in. It is well known that for a wide variety of materials, including insulating materials such as glass or clay \cite{lyklemareview,saini,lobbus,sonnefeldsilica,obrianclay,leroyclay}, the charges in the Stern layer are not static. It has even been previously shown that the mobility of the Stern-layer charges is comparable to the mobility of ions in bulk \cite{lyklemareview,minormob,lobbusmob}. 

In the most common case, the surface charge in the Stern layer originates from a chemical reaction with dissolved ions in the fluid, either via an adsorption or a desorption reaction. The surface charge is therefore not a fixed quantity, but is determined by a charge regulation process \cite{chargeregulation}. In this work we will consider the desorption reaction SC$\rightleftharpoons$ S$^-$+C$^+$, where a neutral surface group SC releases a cationic counter ion C$^+$ leaving behind a charged, covalently bound surface group S$^-$. This reaction represents, for example, a deprotonation reaction if we identify C$^+$ as a proton. In equilibrium, the balance of this reaction is given by the Langmuir desorption isotherm $f=(1+\rho_{\rm C,s}/K)^{-1}$ \cite{chargeregulation,CRChan,CRBorkovec}, with $f$ the fraction of charged surface groups, $K$ the chemical equilibrium constant, and $\rho_{\rm C,s}$ the counter ion density at the surface. However, the Langmuir isotherm assumes chemical equilibrium at all times, as in previous theories of charge regulation in electrokinetic systems \cite{zukoski,mangelsdorf}, and therefore does not take finite chemical reaction rates into account. We have recently shown, however, that the combination of finite chemical rates with a non-zero Stern-layer mobility leads to novel electrokinetic properties \cite{werkhoven}. In particular, our numerical solutions showed that a lateral fluid flow induces a heterogeneous surface charge on a chemically homogeneous, finite surface, which furthermore provided a first-principles explanation for the experimentally observed influence of fluid flow on the surface chemistry \cite{lis}. In section \ref{sec:PNP} of this paper, we linearize the governing equations used in Ref. \cite{werkhoven}, and solve them analytically with the aid of several simplifying approximations. The analytical solutions exhibit qualitative agreement with the numerical results, but due to the underlying approximations we do not obtain full quantitative agreement. Nevertheless, our analytical approach allows us to identify three important length and associated time scales, summarised in section \ref{sec:times}, as well as three distinct and qualitatively different regimes. For fast reaction rates, the reaction-dominated regime discussed in section \ref{sec:HS}, we reproduce a constant surface charge and the Helmholtz-Smoluchowski equation (\ref{eq:HSE}), while for finite reaction rates we obtain either a linear surface charge profile (diffusion-dominated regime, section \ref{sec:micro}), or a nonlinear profile (conduction-dominated regime, section \ref{sec:reaclim}). Our analysis allows one to categorise all electrokinetic setups in three regimes, such that the characteristics of the surface charge distribution can be predicted or tuned.

\section{The (linearised) Poisson-Nernst-Planck equations}\label{sec:PNP}

We consider an electrokinetic system consisting of two water reservoirs connected by a rectangular channel with height $H$ and length $2L$. We denote the normal and lateral Cartesian coordinate by $z\in [0,H]$ and $x\in [-L,L]$, respectively, and assume translational invariance in the $y$ direction. The top ($z=H$) and bottom ($z=0$) surface of the channel carry chargeable surface groups, which for simplicity we assume to be equal such that the plane at $z=\frac{1}{2}H$ is a symmetry plane of the system. Without loss of generality, we take the fluid flow in the positive $x$ direction. The reservoirs contain three different ionic species labelled by $i=+,-,C$, with valency $z_+=-z_-=1$ and $z_C$. Charge neutrality in the bulk demands $\sum_iz_i\rho_{b,i}=0$, with $\rho_{b,i}$ the bulk concentration of ion $i$ in the two reservoirs, i.e. we do not consider diffusio-osmotic processes here. This fixes the Debye screening length, as $\lambda_D=\sqrt{\epsilon k_{\rm B}T/(e^2\sum_iz_i^2\rho_{b,i})}$, the typical thickness of the EDL, with $k_{\rm B}$ the Boltzmann constant, $T$ the temperature, and $e$ the elementary charge. 

We denote the position and time dependent concentration and flux of the three ion species by $\rho_i({\bf r},t)$ and ${\bf J}_i({\bf r},t)$, respectively, the electric potential by $\psi({\bf r},t)$ and the fluid velocity and pressure by ${\bf u}({\bf r},t)$ and $p({\bf r},t)$, respectively. These quantities are governed by the Poisson-Nernst-Planck-Navier-Stokes (PNP-NS) equations \cite{hunter},
\begin{gather}
\begin{aligned}
&\dfrac{\partial \rho_i}{\partial t}=-\nabla\cdot  {\bf J}_i; \hspace{3mm} {\bf J}_i =-D\left(\nabla\rho_i+\frac{ez_i\rho_i}{k_{\rm B}T}\nabla\psi\right)+\rho_i{\bf u};\\
&m\dfrac{\partial {\bf u}}{\partial t}=-m({\bf u} \cdot\nabla){\bf u}-\nabla p+\eta\nabla^2{\bf u}-\sum\limits_i z_ie\rho_i \nabla \psi;\\
&\nabla\cdot {\bf u}=0; \qquad \nabla^2\psi=-\frac{e}{\epsilon}\sum\limits_iz_i\rho_i, 
\end{aligned}
\label{eq:equations}
\raisetag{50pt}
\end{gather}
where $D$ is the diffusion constant (assumed for simplicity to be equal for all ionic species), and $m$ the mass density of the fluid. The PNP-NS equations combine the Poisson equation for the electrostatic potential, the incompressible Navier-Stokes equation for the fluid flow, the Nernst-Planck equation for ionic transport and the continuity equation for the ion densities.

These equations are then to be coupled to a dynamic Stern layer. In our theoretical framework, we treat the density of surface charges $\sigma(x,t)$, a 2D analogue of $\rho_i({\bf r},t)$, as a dynamic variable. The surface charges are produced by a chemical reaction SC$\rightleftharpoons$ S$^-$+C$^+$, and therefore $\sigma(x,t)$ is not necessarily a locally conserved quantity. However, since the total number of counter ions must be conserved, the production rate $R$ of surface charges is equal to the counter-ion flux leaving the surface ${\bf n}_s\cdot{\bf J}_{C,s}$, with ${\bf n}_s$ the normal vector of the solid-liquid interface pointing into the liquid. Denoting the flux of surface charges in the $x$-direction  by $j_{\sigma}(x,t)$, the 2D analogue of ${\bf J}_i({\bf r},t)$, we can write the continuity equation for $\sigma(x,t)$ as
\begin{equation}\label{eq:surfcont}
\frac{\partial \sigma}{\partial t}=-\frac{\partial j_{\sigma}}{\partial x}+{\bf n}_s\cdot {\bf J}_{C,s}.
\end{equation}
The production rate of surface charges is governed by the chemical rate equations, where the surface charge production (annihilation) rate is proportional to the density of uncharged (charged) sites. Furthermore, we assume a Nernst-Planck-like equation for $j_{\sigma}$, derived via a dynamical density functional theory in Appendix \ref{sec:DFT}, with the noticeable absence of convection in the Stern layer. The resulting equations that govern the surface dynamics, and therefore can be seen as the boundary conditions to the PNP-NS equations (\ref{eq:equations}), are given by
\begin{align}
{\bf n}_s\cdot \nabla \psi_s&=-\frac{z_{\sigma}e\sigma}{\epsilon}; \qquad {\bf u}_s=0;\label{eq:bc1}\\
j_{\sigma}&=-D_s\left(\frac{1}{1-\sigma/\Gamma}\frac{\partial\sigma}{\partial x}+z_{\sigma}\frac{e\sigma}{k_{\rm B}T}\frac{\partial \psi_s}{\partial x}\right);\label{eq:bc2}\\
{\bf n_s\cdot J_{\rm C,s}}=R&=k^{\rm des}(\Gamma-\sigma)-k^{\rm ads}\sigma\rho_{\rm C,s},\label{eq:coupling}
\end{align}
with $D_s$ is the diffusion constant of the surface charges, $z_{\sigma}=-z_C$ the valency of the surface charges, $\Gamma$ the density of surface sites, and $k^{\rm ads}$ and $k^{\rm des}$ the adsorption and desorption rate constants, respectively. Here, and throughout this work, we use the subscript ``s" to denote surface quantities, such that ${\bf u}_s=0$ enforces a no-slip boundary condition and $\rho_{C,s}(x)\equiv \rho_{C}(x,z=0)$ the counter ion concentration at the surface. The diffusion contribution ($\propto \partial_x\sigma$) to $j_{\sigma}$ must in general be adjusted by a factor $(1-\sigma/\Gamma)^{-1}$ as the surface groups cannot be multiply occupied \cite{lattice}. Furthermore, we assume that $D_s$ does not depend the surface concentration $\sigma$. For high concentrations, however, the diffusion constant depends non-trivially on the concentrtaion \cite{lattice2}. However, since at most a few percent of the total number of sites is charged, $\sigma$ is sufficiently small to safely assume $D_s$ to be constant. Similarly, we have assumed that the bulk diffusion constant $D$ in Eq. (\ref{eq:equations}) is homogeneous throughout the system, while in general this depends on for example the distance to the surface \cite{difconst}. For simplicity, we leave out these higher order effects. In equilibrium all fluxes vanish, and for ${\bf J}_C=0$ we recover from Eq. (\ref{eq:coupling}) the standard Langmuir desorption equation $\sigma=\Gamma/(1+\rho_{C,s}/K)$, with $K=k^{\rm des}/k^{\rm ads}$ equal to the chemical equilibrium constant of the reaction SC$\rightleftharpoons$ S$^-$+C$^+$.

The governing equations (\ref{eq:equations})-(\ref{eq:coupling}) cannot be solved analytically in general. In this article, however, we will show how to obtain approximate solutions to these equations. As is common in pressure-driven electrokinetic systems, we neglect not only the inertial terms in the Navier-Stokes equation (low Reynolds number, effectively $m=0$), but also the electric body forces on the fluid \cite{hunter}, such that we can ignore the final term on the right hand side of the Navier-Stokes equation (\ref{eq:equations}). The latter approximation can be justified for setups with $H\gg\lambda_D$, and by realising that the body force is localised in the EDL while the pressure gradient extends over the entire channel height. This approximation is confirmed by our numerical calculations (see Appendix \ref{sec:EBF}), which show that including the electric body force has no significant effect on the steady state surface charge profile. The Navier-Stokes equation then reduces to the Stokes equation, and is now decoupled from other quantities. This allows us to solve for ${\bf u}$ and $p$, resulting for an applied pressure drop $\Delta p$ to the standard Poisseuille flow,
\begin{equation}\label{eq:pois}
{\bf u}(z)=-\frac{\partial p}{\partial x}\frac{z(H-z)}{2\eta}{\bf \hat{x}},
\end{equation}
with ${\bf \hat{x}}$ the unit vector in the $x$ direction and $\frac{\partial p}{\partial x}=- \frac{\Delta p}{2L}$. The typical flow velocity can be estimated from Eq. (\ref{eq:pois}). For pressure drops $\Delta p$ no larger than 1 bar, $\eta\sim 1\,$ mPa s, and channel dimensions $H\sim 1\, \mu$m and $L\sim 10\,\mu$m, one arrives at $u_x(\lambda_D)<10^{-2}$ m/s for physically relevant salt concentrations in water.

In the following analysis we will focus solely on the net charge density $\rho_e({\bf r},t)=\sum_iz_i\rho_i({\bf r},t)$ and electric current ${\bf J}_e({\bf r},t)=\sum_iz_i{\bf J}_i({\bf r},t)$. The latter can be written as
\begin{equation}\label{eq:Je}
{\bf J}_e=-D\left(\nabla\rho_e+\frac{e\rho}{k_{\rm B}T}\nabla\psi\right)+\rho_e{\bf u},
\end{equation}
with $\rho({\bf r},t)=\sum_iz_i^2\rho_i({\bf r},t)$ the total local ionic strength. As is common in a linearised theory of electrokinetic systems, we assume here that $\rho$ is constant throughout the system and equal to its bulk value,  $\rho=\sum_iz_i^2\rho_{i,b}$ . At steady state, $\nabla\cdot{\bf J}_e=0$, together with the use of the Poisson equation to eliminate the electric potential in favour of the charge density, we obtain the governing equation for $\rho_e(x,z)$,
 \begin{equation}\label{eq:rhoe}
\left(-D\frac{\partial^2}{\partial x^2}-D\frac{\partial^2}{\partial z^2}+\frac{D}{\lambda_D^2}+ u_x(z) \frac{\partial}{\partial x}\right) \rho_e(x,z)=0.
\end{equation}
We can simplify Eq. (\ref{eq:rhoe}) using scaling arguments. From our previous work, we know that the fluid flow will induce heterogeneities in the $x$ direction, and therefore we estimate that $\partial/\partial x \sim 1/L$. We also know that $\rho_e$ reduces quickly to 0 within a few $\lambda_D$ in the $z$ direction, and hence $\partial/\partial z\sim 1/\lambda_D$. This allows us to define a few characteristic time scales for the EDL,
\begin{equation}
\tau_{\rm L}=\frac{L^2}{D}; \qquad \tau_{\rm EDL}=\frac{\lambda_D^2}{D};\qquad \tau_{\rm adv}=\frac{L}{u_x(\lambda_D)}.
\end{equation} 
Here, $\tau_L$ is the characteristic time for an ion in bulk to diffuse the lateral length $L$, and is of the same order of magnitude as the first term of Eq. (\ref{eq:rhoe}). Additionally, $\tau_{\rm EDL}$ is the characteristic equilibration time of an EDL, and is of the same order of magnitude as the second and third term of Eq. (\ref{eq:rhoe}). Lastly, $\tau_{\rm adv}$ is the characteristic time it takes for an ion in the EDL to be advectivelly transported from one end of the channel to the other. Since $\rho_e$ is only non-zero in the EDL, we evaluated $u_x$ at $z=\lambda_D$. For our geometry, $L\gg\lambda_D$, and thus we can conclude that $\tau_{\rm EDL}\ll\tau_L$, meaning that the diffusion in the lateral direction (first term Eq. (\ref{eq:rhoe})) is negligible compared to diffusion in the normal direction (second term Eq. (\ref{eq:rhoe})). Moreover, since $D/\lambda_D\sim 0.1 $m/s$>u_x(\lambda_D)$ for our parameter choice of interest, we have  $\tau_{\rm EDL}\ll \tau_{\rm adv}$ implying that convection in the lateral direction (last term Eq. (\ref{eq:rhoe})) is negligible with respect to diffusion in the normal direction. After plugging in typical quantities, one indeed finds that $\tau_{\rm EDL}$ is of the order of (tens of) nanoseconds, while $\tau_{\rm adv}$ is of the order of miliseconds or larger. Eq. (\ref{eq:rhoe}) now reduces to a simple differential equation,
\begin{equation}\label{eq:EDL}
\frac{\partial^2\rho_e}{\partial z^2}=\frac{1}{\lambda_D^2}\rho_e \qquad \Rightarrow \qquad \rho_e(x,z)=-\zeta(x)\rho e^{-z/\lambda_D},
\end{equation}
where $\zeta(x)$ is an integration constant that remains to be found. As we will show below, we can identify $\zeta(x)=\beta e(\psi(x,0)-\psi(x,\frac{1}{2}H))$ as the (heterogeneous) dimensionless zeta potential at steady state. The second integration constant has been set to zero since $\rho_e(x,z\rightarrow\frac{1}{2}H)=0$ from $H\gg\lambda_D$. Eq. (\ref{eq:EDL}) is analogous to the equilibrium linear Poisson-Boltzmann equation for the charge density, and is a direct consequence of $\tau_{\rm EDL}\ll \tau_{\rm adv}$: the EDL is equilibrated in the $z$ direction as convection is typically not strong enough to deform the EDL significantly.

To determine the integration constant $\zeta(x)$, we apply the boundary conditions Eqs. (\ref{eq:surfcont}) and (\ref{eq:bc1})-(\ref{eq:coupling}). To facilitate the calculations, we make use of the fact that for the majority of surfaces only a small fraction of the sites are charged, $\sigma\ll \Gamma$. For instance, only a few percent of the surface groups of silica are charged under typical conditions ($3<$pH$<11$, 1 mM$<\rho_s<$100 mM)  \cite{silica}. With this assumption, the equation for $\sigma(x)$ simplifies to
\begin{equation}\label{eq:sigma}
D_s\left(\frac{\partial^2\sigma}{\partial x^2} +\frac{z_{\sigma}e}{k_{\rm B}T}\frac{\partial}{\partial x}\left(\sigma\frac{\partial\psi_s}{\partial x}\right)\right)+k^{\rm des}\Gamma-\left(k^{\rm des}+k^{\rm ads}\rho_{C,s}\right)\sigma=0.
\end{equation}
Eq. (\ref{eq:sigma}) constitutes a diffusion-conduction-reaction problem coupled to the 3D-channel via the in-plane electric field $\frac{\partial \psi_s}{\partial x}$ and the counter ion density $\rho_{C,s}$.  As a consequence, three regimes will arise depending on which process is dominant. This is reminiscent of a convection-diffusion problem \cite{convdif} with a linear/exponential density profile for diffusion/convection dominated systems. In our case, the role of convection in the Stern layer is played by conduction, but we will analogously find a linear/exponential in the diffusion/conduction limited regime.

Since the fluid flow is in the positive $x$ direction, the streaming electric field $-\partial\psi_s/\partial x$ must have the same sign as the surface charge such that no net charge is transported between the two reservoirs. It is convenient to separate the sign and magnitude of the streaming electric field. Thus, we define $-\beta e\partial\psi_s/\partial x=z_{\sigma} E$, where $E$ is a positive quantity with dimensions of inverse length. To solve for $\sigma(x)$, we further assume that both $\rho_{C,s}$ and $E$ are spatially constant. While this is a valid assumption in simple electrokinetic systems, we have shown recently \cite{werkhoven} that such approximations are no longer valid when both Stern-layer conduction and finite chemical rates are taken into account; we found that a heterogeneous surface charge leads to a heterogeneous streaming electric field and counter ion density along the surface. Nevertheless, approximating $E$ and $\rho_{C,s}$ to be spatially constant allows us to solve for $\sigma(x)$ and determine $\rho_e$, ${\bf J}_e$ and $E$. In principle one could then reinsert the solutions in Eq. (\ref{eq:sigma}) and obtain an improved $\sigma(x)$, $\rho_{C,s}$ and $E$. However, in this work we will refrain from applying an iterative scheme and aim for an analytical and qualitative understanding of the electrokinetic phenomena.

For a spatially constant $E$ and $\rho_{C,s}$, Eq. (\ref{eq:sigma}) is straightforward to solve. Since $z_{\sigma}^2=1$, we find
\begin{equation}\label{eq:sigmasol}
\sigma(x)=\sigma_{\rm eq}\left(1+a_+ e^{k_+x}+a_-e^{k_-x}\right),
\end{equation}
with $a_{\pm}$ integration constants and $\sigma_{\rm eq}\equiv\Gamma(1+\frac{\rho_{\rm C,s}}{K})^{-1}$ the equilibrium surface charge density. The wavenumbers $k_{\pm}=\frac{1}{2}E\pm\frac{1}{2}\sqrt{E^2+4\lambda_{\rm reac}^{-2}}$ set the relevant lateral length scales, with $\lambda_{\rm reac}\equiv \sqrt{D_s\tau_{\rm reac}}$ (discussed in more detail in section \ref{sec:times}), and $\tau_{\rm reac}=\left(k^{\rm des}+k^{\rm ads}\rho_{C,s}\right)^{-1}$ the characteristic time scale of the chemical reaction. The amplitudes $a_{\pm}$ can be determined by imposing that the surface current vanishes at the end-points of the charged surface, $j_{\sigma}(\pm L)=(-D_s\partial_x\sigma+D_sE\sigma)|_{x=\pm L}=0$. The integration constants $a_{\pm}$ are then found to be
\begin{equation}
a_{\pm}=\frac{E}{k_{\mp}}\frac{\sinh k_{\mp}L}{\sinh(k_{\pm}-k_{\mp})L}.
\end{equation}
Note that $\sigma(x)$ does not depend on $z_{\sigma}$, and that in equilibrium, $E\rightarrow 0$, and hence $\sigma(x) \rightarrow \sigma_{\rm eq}$. For non-zero $E$, we find a double exponential profile. This reduces to either a linear profile for $k_{\pm}L\ll 1$, or a single exponential profile if $k_{\pm}L\gg 1$.

The final unknown, the integration constant $\zeta(x)$, can be determined using Eq. (\ref{eq:coupling}), coupling $\sigma(x)$ to $\rho_e(x,0)$, see Appendix \ref{sec:calculations}. To facilitate this calculation, we rewrite the counter ion flux as $\hat{z}\cdot{\bf J}_C(x,0)=z_C\hat{z}\cdot{\bf J}_e(x,0)$,
where we have used ${\bf n}_s=\hat{z}$, $\hat{z}\cdot{\bf J}_+(x,0)=\hat{z}\cdot{\bf J}_-(x,0)=0$. Using the solutions to $\rho_e$, Eq. (\ref{eq:EDL}), and $\sigma$, Eq. (\ref{eq:sigmasol}), we find
\begin{equation}\label{eq:zeta}
\zeta(x)=\zeta_{\rm eq}\left(1+a_+ e^{k_+x}+a_-e^{k_-x}\right).
\end{equation}
Here, we identified the dimensionless equilibrium zeta potential $\zeta_{\rm eq}=z_{\sigma}4\pi\lambda_B\lambda_D\sigma_{\rm eq}$ from linear Poisson-Boltzmann theory.  Comparing Eq (\ref{eq:sigmasol}) with Eq. (\ref{eq:zeta}), we see that $\zeta(x)$ is proportional to the steady state surface charge, and indeed can be interpreted $\zeta(x)$ as the steady state zeta potential.

To determine $E$, we impose that at any position $x$ no net current passes through any channel slice with normal ${\bf \hat{x}}$. This condition is a direct consequence of the vanishing divergence of ${\bf J}_e$ and the open circuit geometry, hence
\begin{equation}\label{eq:netcurrent}
z_{\sigma}j_{\sigma}(x)+\int_0^{\frac{1}{2}H} \dif z J_{e,x}(x,z)=0,
\end{equation}
where $z_{\sigma}j_{\sigma}(x)$ is the net charge current through the Stern layer, determined using Eqs. (\ref{eq:sigmasol}) and (\ref{eq:bc2}), and due to symmetry we only integrate over half the channel height. Eq. (\ref{eq:netcurrent}) is a local condition, and we will in general find $E$ to depend on $x$ (see below). This, however, contradicts our previous assumption that $E$ is spatially constant, and it is at this point that our analytic approach is inconsistent. Nevertheless, the results of this analytic approach allow us obtain approximate solution, and agree qualitatively well with the full numerical solutions. Despite the inconsistency, this approach gives us physical insight in the system, and allows us to identify three separate regimes.

\section{Three electrokinetic regimes}

\subsection{Governing time and length scales}\label{sec:times}

We identify two physically important length scales, that appear in the definition of the wavenumbers $k_{\pm}$, 
\begin{equation}\label{eq:lengths}
\lambda_{\rm reac}=\sqrt{D_s\tau_{\rm reac}};\qquad \lambda_{\rm cond}=\frac{1}{E}.
\end{equation}
We can interpret the first length scale $\lambda_{\rm reac}$ as the typical distance a surface charge traverses diffusively during a time $\tau_{\rm reac}$, that is, the typical distance travelled between ad- and desorption. The conductive length scale $\lambda_{\rm cond}$ can be interpreted as the distance a monovalent ion needs to travel in order to gain an energy equal to $k_{\rm B}T$ due to the streaming electric field. The dynamics of the system is fully determined by $\lambda_{\rm reac}$, $\lambda_{\rm cond}$, and the channel length $L$.

Alternatively, we can identify an equivalent time scale for each length scale using the surface diffusion constant $D_s$. Eq. (\ref{eq:lengths}) shows that $\tau_{\rm reac}$ is the equivalent time scale of $\lambda_{\rm reac}$. By introducing a conductive velocity $v_{\rm cond}=D_sE$ and a diffusive velocity $v_{\rm dif}=D_s/L$, we can transform $L$ and $\lambda_{\rm cond}$ into equivalent time scales,
\begin{equation}\label{eq:times}
\tau_{\rm dif}=\frac{L}{v_{\rm dif}}=\frac{L^2}{D_s}\qquad \tau_{\rm cond}=\frac{\lambda_{\rm cond}}{v_{\rm cond}}=\frac{1}{D_sE^2}.
\end{equation}
We can interpret $\tau_{\rm dif}$ as the characteristic time for a Stern-layer charge to diffuse across the channel length, and $\tau_{\rm cond}$ as the characteristic time after which a Stern-layer charge has gained one thermal energy unit due to the streaming electric field. Together with $\tau_{\rm reac}$, the three time scales can be used equivalently to the three length scales to characterise the electrokinetic system, as we can express the ratio between every pair of length scales as the ratio between the two equivalent time scales:
\begin{equation}
\frac{\lambda_{\rm reac}}{L}=\sqrt{\frac{\tau_{\rm reac}}{\tau_{\rm dif}}}; \qquad \frac{L}{\lambda_{\rm cond}}=\sqrt{\frac{\tau_{\rm dif}}{\tau_{\rm cond}}}; \qquad \frac{\lambda_{\rm cond}}{\lambda_{\rm reac}}=\sqrt{\frac{\tau_{\rm cond}}{\tau_{\rm reac}}}.
\end{equation}
The three distinct characteristic times allow us to identify three electrokinetic regimes, defined by the smallest time (or associated length). While the three length scales appear naturally in the analytical description, we found it more intuitive to consider the three time scales when considering the different dynamical regimes. In the reaction-dominated regime, characterised by the near-equilibrium of the adsorption/desorption process, $\tau_{\rm reac}\ll \tau_{\rm cond},\tau_{\rm dif}$ (and hence $\lambda_{\rm reac}\ll L, \lambda_{\rm cond}$), to be discussed in section \ref{sec:HS}, we obtain the standard Helmholtz-Smoluchowski picture with a constant surface charge and electric field, except for a region of size $\lambda_{\rm reac}$ around the edges at $x=\pm L$. In the diffusion-dominated regime, $\tau_{\rm dif}\ll \tau_{\rm cond},\tau_{\rm reac}$ ($L\ll \lambda_{\rm cond},\lambda_{\rm reac}$), discussed in section \ref{sec:micro}, the surface charge is heterogeneous over the entire surface and linear in the lateral position $x$. Consequently, the streaming electric field $E$ is also heterogeneous, but we find that the streaming potential approximately conforms to the Helmholtz-Smoluchowski equation Eq. (\ref{eq:HSE}). However, in the conduction-dominated regime $\tau_{\rm cond}\ll \tau_{\rm reac},\tau_{\rm dif}$ (and hence $\lambda_{\rm cond}\ll \lambda_{\rm reac},L$), to be discussed in section \ref{sec:reaclim}, Eq. (\ref{eq:HSE}) no longer holds, and both $\sigma$ and $E$ are heterogeneous and nonlinear function of $x$.  

\subsection{Reaction-dominated regime}\label{sec:HS}

\begin{figure}[!ht]
\centering
\begin{subfigure}{.5\textwidth}
  \centering
  \caption{}
  \includegraphics[width=0.95\textwidth]{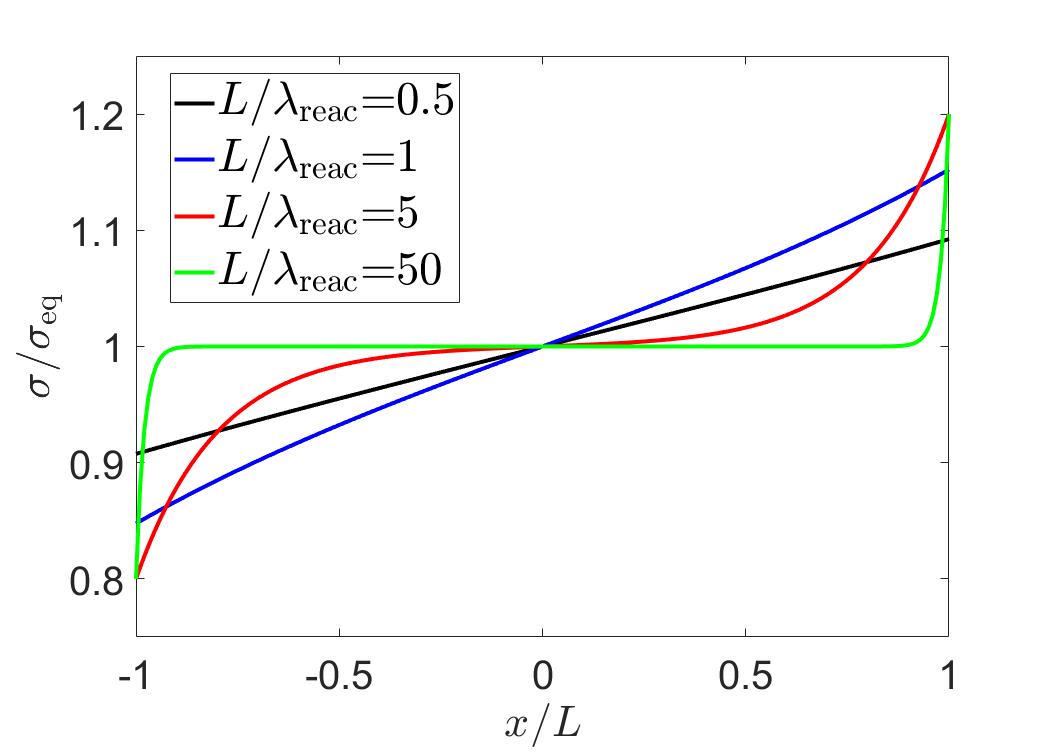}
  \label{fig:Chargelin}
\end{subfigure}%
\begin{subfigure}{.5\textwidth}
  \centering
  \caption{}
  \includegraphics[width=0.95\textwidth]{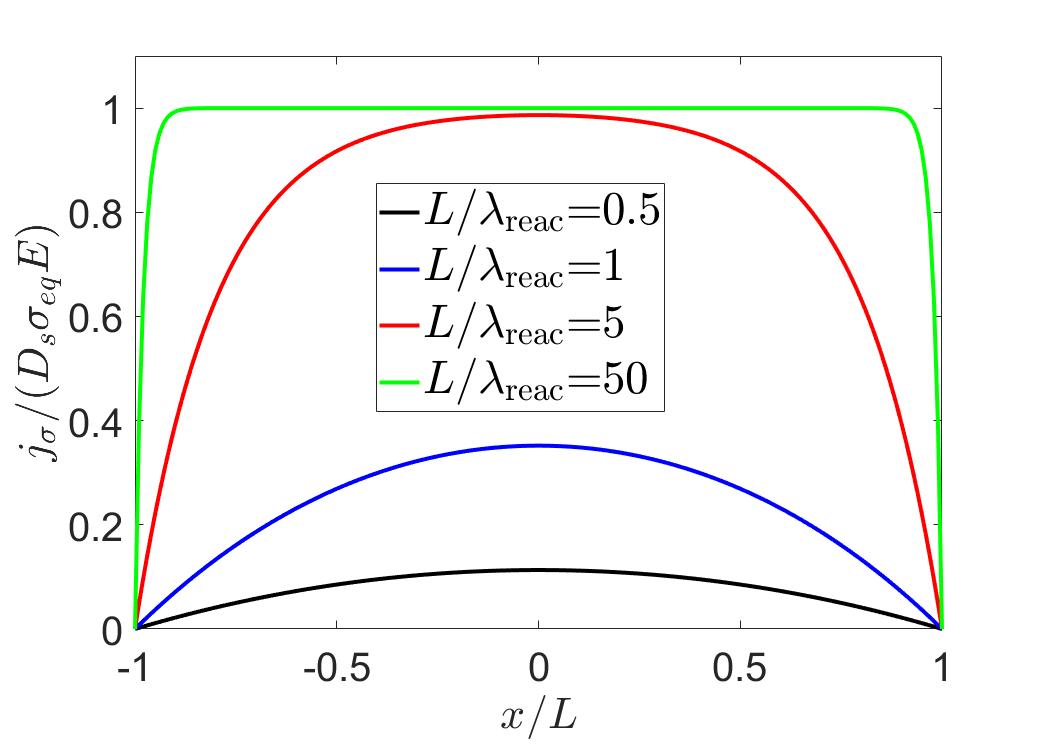}
  \label{fig:jslin}
\end{subfigure}
\caption{Surface charge density $\sigma(x)$ (a) and Stern-layer current $j_{\sigma}(x)$ (b) between channel inlet ($x=-L$) and outlet ($x=L$), in the limit $\lambda_{\rm reac}\ll\lambda_{\rm cond}$ according to Eqs. (\ref{eq:surflin1}) and (\ref{eq:surflin2}) for varying values of $L/\lambda_{\rm reac}$. For (a) we used $\lambda_{\rm reac}/\lambda_{\rm cond}=0.2$ ($j_{\sigma}$ does not depend on this quantity). As $L/\lambda_{\rm reac}$ decreases, the system leaves the reaction-dominated regime and enters the diffusion-dominated regime (for which $\lambda_{\rm reac}\ll\lambda_{\rm cond}$ holds) and $\sigma$ and $j_{\sigma}$ become increasingly heterogeneous.}
\label{fig:HSESolutions}
\end{figure}

In the first regime, we consider systems where the chemical reaction rates are the fastest process in the system. In this regime, therefore, we expect to find a constant surface charge and consequently the standard Helmholtz-Smoluchowski equation (\ref{eq:HSE}). In terms of time scales we have $\tau_{\rm reac}\ll\tau_{\rm dif},\tau_{\rm cond}$, which implies that $\lambda_{\rm reac}$ is the smallest length scale, i.e. $\lambda_{\rm reac}\ll L,\lambda_{\rm cond}$. It should be noted that, since $\tau_{\rm cond},\tau_{\rm dif}\propto D_s^{-1}$, a system without Stern-layer conduction ($D_s=0$) cannot be diffusion- or conduction-dominated and is in fact always in the reaction-dominated regime. The resulting equations in the reaction-dominated regime do not depend on the ratio between $\lambda_{\rm cond}$ and $L$, which we can therefore leave unspecified. In this limit, the wavenumbers can be approximated by  $k_+=-k_-=\lambda_{\rm reac}^{-1}$. This simplifies the solution for $\sigma(x)$, Eq. (\ref{eq:sigma}), the Stern-layer flux $j_{\sigma}(x)$, Eq. (\ref{eq:bc2}) (see Appendix \ref{sec:calculations} for the general expression), and the surface charge production rate $R(x)=-{\bf n}_s\cdot {\bf J}_C$. Eq. (\ref{eq:coupling}), as
\begin{align}
\sigma(x)&=\sigma_{\rm eq}\left(1+E\lambda_{\rm reac}\frac{\sinh x/\lambda_{\rm reac}}{\cosh L/\lambda_{\rm reac}}\right);\label{eq:surflin1}\\
j_{\sigma}(x)&=D_s\sigma_{eq}E\left(1-\frac{\cosh x/\lambda_{\rm reac}-E\lambda_{\rm reac}\sinh x/\lambda_{\rm reac}}{\cosh L/\lambda_{\rm reac}}\right)\approx D_s\sigma_{\rm eq}E\left(1-\frac{\cosh x/\lambda_{\rm reac}}{\cosh L/\lambda_{\rm reac}}\right);\label{eq:surflin2}\\
R(x)&=\frac{\sigma_{\rm eq}E\lambda_{\rm reac}}{\tau_{\rm reac}}\frac{\sinh x/\lambda_{\rm reac}}{ \cosh L/\lambda_{\rm reac}}.\label{eq:surflin3}
\end{align}
In Figs. \ref{fig:HSESolutions}(\subref{fig:Chargelin}) and \ref{fig:HSESolutions}(\subref{fig:jslin}) we plot $\sigma(x)$ and $j_{\sigma}(x)$ respectively, for several values of $L/\lambda_{\rm reac}$ smaller and larger than unity. From $\lambda_{\rm reac}\ll L$, we can deduce from Eqs. (\ref{eq:surflin1})-(\ref{eq:surflin3}) that $\sigma(x)\approx \sigma_{eq}$, $j_{\sigma}(x)\approx D_s\sigma_{eq}E$ and $R\approx 0$ for all $x$ except within a few $\lambda_{\rm reac}$ away from $x=\pm L$. In Fig. \ref{fig:HSESolutions}, we show this deviation near $x=\pm L$ for both $\sigma(x)$ and $j_{\sigma}(x)$, which is claerly visible for several values of $L/\lambda_{\rm reac}$. This edge effect is a direct result of the boundary condition $j_{\sigma}(\pm L)=0$, that stems from the fact that our surface has finite length. In order for a non-zero $j_{\sigma}$ to develop, counter ions must adsorb at the inlet and desorb at the outlet, which is only possible if $\sigma$ deviates from $\sigma_{\rm eq}$. This explains the heterogeneities of $\sigma$ shown in Fig. \ref{fig:HSESolutions} that persist even for large $L/\lambda_{\rm reac}$.  Even in the classical Helmholtz-Smoluchowski setting, a finite Stern-layer conduction implies that at the edges of the surface $\sigma$ deviates from its equilibrium value. The range of this inhomogeneity is given by $\lambda_{\rm reac}$, as can be seen in Fig. \ref{fig:HSESolutions}. Consequently, the Stern-layer current and surface charge profile are constant up to a few $\lambda_{\rm reac}$ from the edges of the surface. The amplitude of the relative deviation is interestingly given by $\lambda_{\rm reac}/\lambda_{\rm cond}$. This edge effect exists purely due to the surface charge discontinuity at $x=\pm L$, but $\lambda_{\rm reac}$ is nevertheless not to be confused with the healing length $\ell=H$Du introduced by Khair and Squires \cite{khairsquires}, which also arises in the absence of Stern-layer conduction. 

Imposing a vanishing net current at every position $x$, according to Eq. (\ref{eq:netcurrent}), we find the streaming electric field $E$,
\begin{equation}\label{eq:EHSregime}
E=\frac{|\partial_x p|\zeta_{\rm eq}\epsilon}{\eta G_b}\left[1+\dfrac{1}{H}\dfrac{D_s\sigma_{eq}}{D\rho}\right]^{-1},
\end{equation}
with $\partial_x p$ a short-hand notation for $\frac{\partial p}{\partial x}$. In this limit, we recover the standard Helmholtz-Smoluchowski equation (\ref{eq:HSE}) with a spatially constant electric field consistent with our assumptions. However, just like $j_{\sigma}$ and $\sigma$, $E$ is not constant close to the edges and Eq. (\ref{eq:EHSregime}) only holds several $\lambda_{\rm reac}$ from $x=\pm L$. The contributions of the edge effect to the streaming potential $\Delta\Phi=\int_{-L}^L{\rm d}x E$ are negligible (note the antisymmetric nature of the edge effect).  Furthermore, we can identify using Eq. (\ref{eq:EHSregime}) the Stern-layer contribution to the surface conduction $G_s=D_s\sigma_{eq}\beta e^2=G^S_s$. The latter is proportional to the charge carrier density $\sigma$ in the Stern layer and a 2D-analogue of $G_b=D\rho\beta e^2$. Note that in our calculations, the EDL surface conductivity $G_s^{EDL}$, which originates from the increased ion density in the EDL, does not appear since we assumed that $\rho$ is spatially constant. 

\subsection{Diffusion-dominated regime}\label{sec:micro}

In the second dynamic regime, the diffusion time $\tau_{\rm dif}$ is the smallest time scale, and the channel length $L$ is the smallest length scale, $L \ll \lambda_{\rm cond},\lambda_{\rm reac}$. Consequently, this implies that the dimensionless streaming potential $\beta e \Delta\Phi\sim EL=L/\lambda_{\rm cond}$ is small. Analogously to the Helmholtz-Smoluchowski regime, the ratio between the other two lengths, $\lambda_{\rm cond}/\lambda_{\rm reac}$, will have no significant impact on the results. We can use Eq. (\ref{eq:surflin1}), which was derived assuming only $\lambda_{\rm reac}\ll\lambda_{\rm cond}$, but now with $L\ll \lambda_{\rm reac}$ to write the surface charge, flux and chemical production rate as
\begin{equation}\label{eq:linfunc}
\begin{aligned}
\sigma(x)&\approx\sigma_{\rm eq}\left(1+Ex\right);\\
j_{\sigma}&\approx \frac{1}{2}D_s\sigma_{\rm eq}E\frac{\tau_{\rm dif}}{\tau_{\rm reac}}(1-\frac{x^2}{L^2});\\
R(x)&\approx \frac{\sigma_{\rm eq}E}{\tau_{reac}}x.
\end{aligned}
\end{equation}
In this parameter regime, we thus recover a linear profile for the surface charge density $\sigma(x)$ also found numerically \cite{werkhoven} and shown in Fig. \ref{fig:HSESolutions}(\subref{fig:Chargelin}). We can intuitively understand this linear profile by realising that the system is diffusion dominated. For a surface with translation invariance in one direction, the steady state would then be given by a linear profile. The liner profile is maintained because the chemical reaction is not fast enough to force $\sigma$ to the equilibrium value ($\tau_{\rm reac}\gg \tau_{\rm dif}$). Additionally, since $j_{\sigma}\propto \tau_{\rm dif}/ \tau_{\rm reac}$, the surface flux is very small, as can be observed in Fig. \ref{fig:HSESolutions}(\subref{fig:jslin}). At steady state, the surface charge profile is therefore determined by a balance between the conduction caused by $E$ and diffusion is the opposite direction, which explains why the slope of $\sigma(x)$ is given by the electric field $E$. This result is analogous to a diffusion-dominated convection-diffusion problem. Within the diffusion-dominated regime, we obtain from Eq. (\ref{eq:netcurrent}) a new expression for the streaming electric field,
\begin{equation}\label{eq:Elin}
z_{\sigma}E(x)=\frac{|\partial_x p|\zeta_{\rm eq}\epsilon}{\eta G_b}\left[1+|\zeta_{\rm eq}|\left(\frac{2\lambda_D}{H}-\frac{|\partial_x p|\epsilon}{\eta G_b}x\right)\right]^{-1}=\frac{z_{\sigma}E^{HS}}{1-E^{HS}x}.
\end{equation}
Here we introduced $E^{HS}=\frac{|\partial_x p||\zeta_{\rm eq}|\epsilon}{\eta G_b}$, the (magnitude of the) streaming electric field as predicted by the Helmholtz-Smoluchowski equation (\ref{eq:HSE}) without Stern-layer conduction. Since we have defined $\zeta_{\rm eq}$ as the dimensionless equilibrium zeta potential, $E^{HS}$ has dimensions of inverse length. Note that the solution does not depend on $D_s$ (except for the restriction that $\tau_{\rm reac}\gg\tau_{\rm dif},\tau_{\rm cond}\propto D_s^{-1}$) due to the vanishing $j_{\sigma}$. The streaming electric field is, similar to $\sigma(x)$, heterogeneous, with a smaller value than $E^{HS}$ at the inlet and a larger value at the outlet. The impact of Stern-layer conduction is thus indirect in this case, and not direct via a charge current 'leaking' through the Stern layer as in the reaction-dominated regime. The Stern-layer conduction now allows for a heterogeneous surface charge to develop, which causes the heterogeneity of the channel and all other results discussed here.

The result in Eq. (\ref{eq:linfunc}) agrees qualitatively with the full numerical solutions, where the electric field and surface charge density are smaller at the inlet and larger at the outlet with respect to $E^{HS}$. Both the analytical and numerical solution are equal to the Helmholtz-Smoluchowski result in the center ($x=0$) of the channel. The analytical solution does, however, overestimate the heterogeneity of $\sigma(x)$ and $E(x)$. The reason for this is that we assumed that the counter ion concentration at the surface, $\rho_{C,s}$, does not depend on the surface charge. In reality, of course, it does, since an increased surface charge attracts more counter ions. The lack of this regulation mechanism explains the overestimation of the surface charge and electric field.

Eq. (\ref{eq:Elin}) additionally allows us to derive an expression for the streaming potential,
\begin{equation}\label{eq:streampot}
\beta e\Delta \Phi=-z_{\sigma}\int^L_{-L}{\rm d}x\frac{E^{HS}}{1-E^{HS}x}=-z_{\sigma}\log\left(\frac{1+E^{HS}L}{1-E^{HS}L}\right)\approx \beta e\Delta \Phi^{HS},
\end{equation}
where $\beta e\Delta\Phi_S^{HS}\equiv -z_{\sigma}2LE^{HS}$ is the streaming potential  predicted by the Helmholtz-Smoluchowski equation (\ref{eq:HSE}). For small $\Delta \Phi$, which, as we have discussed above, is always the case in the diffusion-dominated regime, we find that the streaming potential can be accurately estimated using the standard Helmholtz-Smoluchowski expression, $\Delta \Phi\approx \Delta \Phi^{HS}$. The lateral heterogeneity therefore has no significant effect on $\Delta \Phi_S$. This can be explained by the quasi-antisymmetric profile of $\sigma(x)$, as the streaming potential is a laterally integrated quantity. This, combined with the tendency to measure at the center of the channel, might explain why such lateral heterogeneities have not been observed yet. Note that Eq. (\ref{eq:streampot}) breaks down for $|\Delta\Phi|\rightarrow 1$. However, we are in the regime where $L\ll \lambda_{\rm cond}$, which implies that $\beta e\Delta \Phi \approx 2EL \ll 1$. The streaming potential will therefore never diverge, but the system will change to a new regime as $\Delta \Phi$ increases. Lastly, we can derive a surprisingly simple expression for the surface charge difference between inlet and outlet in the diffusion-dominated regime,
\begin{equation}\label{eq:Dsigma}
\Delta \sigma\equiv \sigma(L)-\sigma(-L)=\int_{-L}^L{\rm d}x \partial_x\sigma(x)\approx \sigma_{\rm eq} \beta e\Delta \Phi\approx \sigma_{\rm eq} \beta e\Delta\Phi^{HS}.
\end{equation}
The streaming potential therefore gives the fractional difference in surface charge between the inlet and outlet, providing a good measure of the heterogeneity of the system. Given that here the streaming potential is approximately equal to the Helmholtz-Smoluchowski result, Eq. (\ref{eq:Dsigma}) gives a priori a measure of the heterogeneity to be expected, although one should keep in mind that in general our results overestimate the actual heterogeneity.

\subsection{The conduction-dominated regime}\label{sec:reaclim}

The third regime, the conduction-dominated regime, is reached when $\tau_{\rm cond}$ ($\lambda_{\rm cond}$) is the smallest time (length) scale, $\tau_{\rm cond}\ll\tau_{\rm dif},\tau_{\rm reac}$ (and hence $\lambda_{\rm cond}\ll L,\lambda_{\rm reac}$), such that the wavenumbers can be approximated as $k_+\approx E_s$ and $k_-\approx 0$. Also here, the ratio of the remaining lengths does not impact the results. This allows us to simplify Eq. (\ref{eq:sigma}) significantly. The surface charge profile is no longer linear or anti-symmetric compared to the equilibrium value $\sigma_{\rm eq}$, but rather exponential, while $j_{\sigma}$ vanishes
\begin{equation}\label{eq:sigmanonlin}
\sigma(x)=\sigma_{\rm eq}\frac{E L}{\sinh E L}e^{E x}, \qquad j_{\sigma}=-D_s\partial_x \sigma+E \sigma=0.
\end{equation}
It should be noticed that $j_{\sigma}$ only vanishes because we assumed a constant $E$. However, we have already seen that this is no longer generally the case, and $j_{\sigma}$ will in fact not exactly vanish in a fully self-consistent analysis, but our analysis does show that $j_{\sigma}$ is small. Our numerical calculations confirm that in this regime, as well as in the diffusion-dominated regime, $j_{\sigma}$ is negligible compared to the bulk fluxes \cite{werkhoven} as the chemical reaction rates are too small for a significant surface flux to develop. Note that Eq. (\ref{eq:sigmanonlin}) shows that the density profile is exponential, which is analogous to a convection-dominated convection-diffusion problem. If we take $L\ll \lambda_{\rm cond}$ ($EL\ll 1$), we recover the same linear profile as in the diffusion-dominated regime Eq. (\ref{eq:linfunc}).

\begin{figure}[!ht]
\centering
\includegraphics[width=0.5\textwidth]{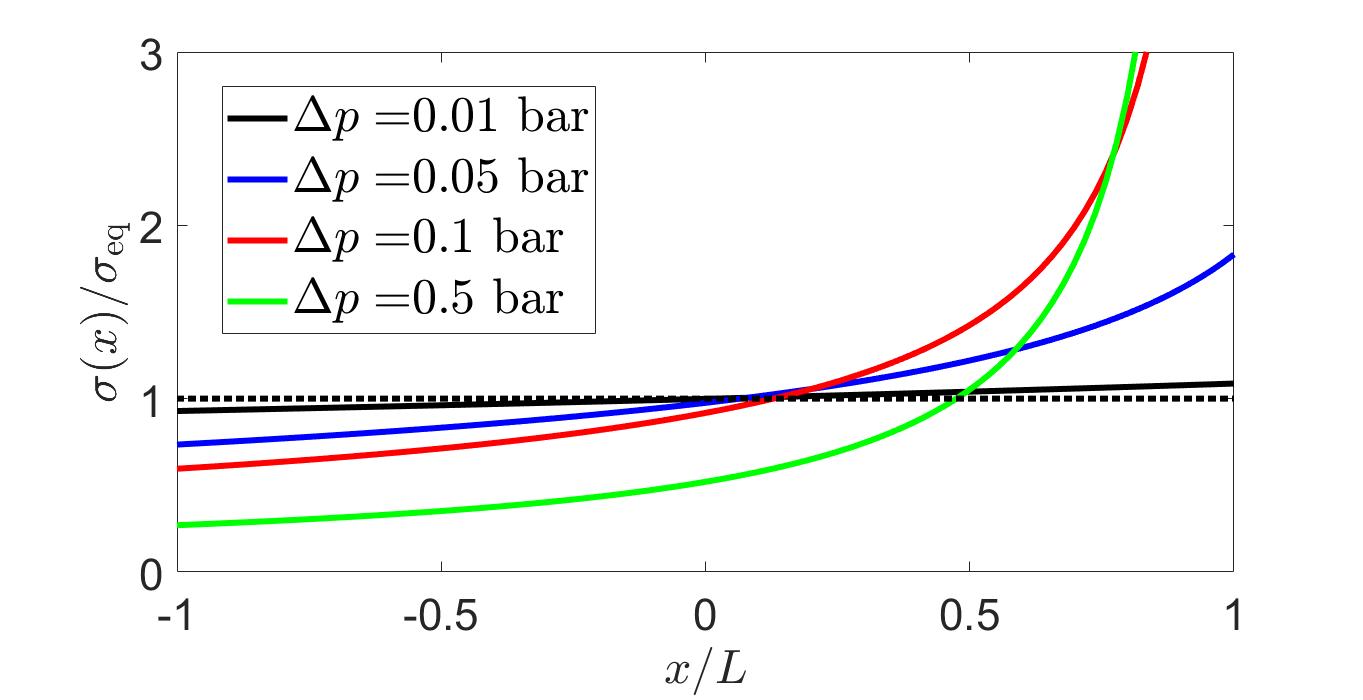}
\caption{Surface charge profile $\sigma(x)$ induced by the electric field $E(x)$, obtained numerically as a solution to Eq. (\ref{eq:jenonlin}), for several values of the pressure drop $\Delta p$. The profiles cross the equilibrium value at $x>0$, and agree qualitatively with the full numerical solutions \cite{werkhoven}.}\label{fig:sc}
\end{figure}

To determine $E$ in the conduction-limited regime we again impose a vanishing net charge current for every $x$, Eq. (\ref{eq:netcurrent}), from which we obtain the condition
\begin{equation}\label{eq:jenonlin}
\zeta_{\rm eq}\frac{EL}{\sinh EL}e^{Ex}\left(2\lambda_DE-\frac{|\partial_xp|H\epsilon}{\eta G_b}\right)+z_{\sigma}EH=0.
\end{equation}
Eq. (\ref{eq:jenonlin}) can be solved numerically for $E(x)$, and we obtain qualitatively similar behaviour as in the full numerical calculations \cite{werkhoven} which is of course inconsistent with our assumptions that $E$ is spatially constant. In Fig. \ref{fig:sc} we plot the resulting surface charge profile according to Eqs. (\ref{eq:sigmanonlin}) and (\ref{eq:jenonlin}) for several values of the pressure drop across the channel.
For small pressure drops, the surface charge profile is roughly linear, corresponding to the diffusion-dominated regime. For larger pressure drops, the profile becomes highly nonlinear, increasing exponentially as $x$ approaches the outlet position at $x=L$. The point where the profile crosses the equilibrium surface charge is, contrary to the diffusion-dominated regime, no longer at $x=0$. Our semi-analytical results agree qualitatively but not quantitatively with the numerical calculations. In particular, the surface charge close to $x=L$ is much larger in the semi-analytical results. The reason for this must again reside in the assumptions that $\rho_{C,s}$ is constant, similar to the diffusion-dominated regime.

There is, in fact, also some qualitative discrepancy between the numerical calculations and the current analysis. In the numerical calculations the average surface charge decreases with an increasing pressure drop, while Fig. \ref{fig:sc} shows that the average surface increases with an increasing pressure drops. The reason for this probably lies again in the missing charge regulation mechanism discussed above, and the exponential increase of $\sigma(x)$ greatly overestimates the average surface charge. Consequently, the predicted streaming potential is also overestimated, and cannot be accurately determined in the current analytical approach. The breakdown of the theory is not unexpected in this respect, as our linearised theory and lack of regulation work best for small driving forces. Lastly, we note that multiplying Eq. (\ref{eq:jenonlin}) with $L$, we see that the solution is given in terms of $EL$ rather than $E$, if $\Delta p$ is fixed. Hence, in this regime, $\sigma$ is invariant under changes in $L$ if $\Delta p$ is fixed, as was in fact also suggested by the numerical solutions \cite{werkhoven}.

\section{Summary \& Conclusion}

\begin{figure}[!ht]
\centering
\includegraphics[width=0.8\textwidth]{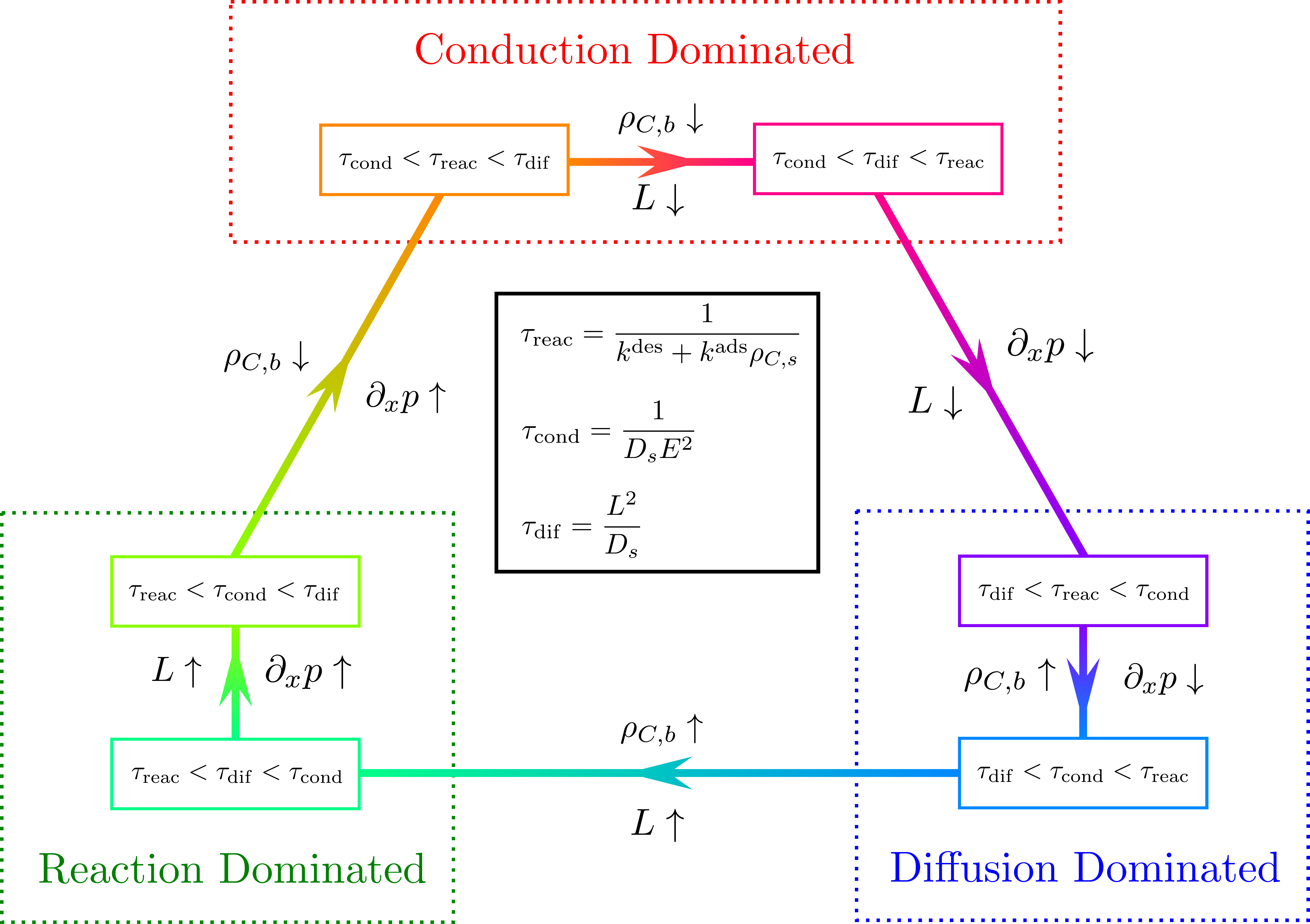}
\caption{Schematic representation of the three regimes and how to transition between them. Each regime is given for each of the 6 possible orderings of the three governing time scales $\tau_{\rm cond}, \tau_{\rm dif}$ and $\tau_{\rm reac}$.}
\label{fig:regimes}
\end{figure}

In this work we revealed some consequences of a chemically and physically dynamic Stern layer on electrokinetic phenomena. By allowing the surface charges to diffuse across the Stern layer and by assigning a finite rate to the adsorption and desorption reactions, we showed that a simple electrokinetic system develops novel properties. We identified three dynamical regimes, schematically represented in Fig. \ref{fig:regimes}. These regimes can be identified via three time scales (or equivalent lengths): the chemical reaction time scale $\tau_{\rm reac}=(k^{\rm des}+k^{\rm ads}\rho_{C,s})^{-1}$, the diffusive time scale $\tau_{\rm dif}=L^2/D_s$ and the conductive time scale $\tau_{\rm cond}=1/(D_sE^2)$. Here, $E=\beta e \partial_x\psi_s$ is the dimensionless streaming electric field and has dimensions of inverse length. The particular regime is determined by which of the three time scales is the smaller one.
There are three basic parameters to change the regime of the electrokinetic system: the pressure gradient $\partial_xp$ changes the conduction time $\tau_{\rm cond}$ and $\lambda_{\rm cond}$, the bulk counter ion density (for example, the pH of the solution) alters $\tau_{\rm reac}$ and $\lambda_{\rm reac}$, while $\tau_{\rm dif}$ varies with $L$. Note that the bulk ionic strength $\rho$ is also an experimentally tunable parameter, which has a similar effect as the pressure gradient; increasing $\rho$ decreases $E$ and thus increases $\tau_{\rm cond}$ and $\lambda_{\rm cond}$. In order to transition between the three regimes, the ordering of the time scales must be changed, which is schematically represented in Fig. \ref{fig:regimes}. 

The first regime, the reaction-dominated regime, emerges when the chemical reaction rates are the fastest process, $\tau_{\rm reac}\ll \tau_{\rm dif},\tau_{\rm cond}$. Due to the high chemical rates we find a mostly constant surface charge profile, except for a region of size $\lambda_{\rm reac}$ away from the ends of the surface. Moreover, we recover the standard Helmholtz-Smoluchowski equation (\ref{eq:HSE}) for the streaming electric field, from which we can read off an explicit expression for the Stern-layer conductivity. The system is in the second regime, the diffusion-dominated regime, if $L$ is the smallest length scale, or equivalently if $\tau_{\rm dif}$ is the smallest time scale. In this regime, the system develops a linear surface charge profile equal to the equilibrium value at the middle of the surface. Although the surface charge and thus the streaming electric field are laterally heterogeneous, the resulting streaming potential is within a good approximation equal to the Helmholtz-Smoluchowski expression Eq. (\ref{eq:HSE}) with zero Stern-layer conductance. For large streaming potential and slow reaction rates, $\tau_{\rm cond}\ll\tau_{\rm dif},\tau_{\rm reac}$, the system reaches the final regime, the conduction-dominated regime. The surface charge profile is then exponential, and is no longer equal to the equilibrium value in the center of the channel. Consequently, the streaming electric field is also exponential, and must be found numerically. In the conduction-dominated regime, the electric field and thus the streaming potential differ significantly from the Helmholtz-Smoluchowski expression. 

We believe that this theoretical framework provides both a deeper physical understanding of the processes at work for a dynamic Stern layer, and is able to predict what properties to expect from an experimental setup. A difficulty is the chemical rates, which seem to be unknown for many surface-electrolyte combinations. Perhaps, by exploring the behaviour of the surface charge by altering the pressure drop or system size, our framework might actually provide insight in the numerical values of the chemical rates.

This work is part of the D-ITP consortium, a program of the Netherlands Organisation for Scientific Research (NWO) that is funded by the Dutch Ministry of Education, Culture and Science (OCW). This work is a part of an Industrial Partnership Program of the Netherlands Organization for Scientific Research (NWO) through FOM Concept agreement FOM-15-0521. Financial support was provided through the Exploratory Research (ExploRe) programme of BP plc.


%

\appendix
\numberwithin{equation}{section}

\section{Governing equations from Dynamic Density Functional Theory}\label{sec:DFT}

The governing equations for the surface charge density $\sigma$ and ion densities $\rho_i$ can be derived using  Dynamical Density Functional Theory (DDFT) \cite{DDFT}, a dynamic extension of Density Functional Theory \cite{DFTbook,DFTEvans}. We consider a system consisting of a 3D bulk region, denoted by $\mathcal{R}$, and a 2D surface, denoted by $\mathcal{S}$ which is part of $\partial \mathcal{R}$, the boundary of $\mathcal{R}$. Here we will denote a position vector in $\mathcal{R}$ with ${\bf r}$ and a position vector in $\mathcal{S}$ with ${\bf r}_s$. The surface charges $\sigma({\bf r}_s)$ are located in $\mathcal{S}$. First we set up the density functional for $\rho({\bf r})$, which consists of an ideal contribution, $\mathcal{F}_{\rm b,id}$, and an excess contribution $\mathcal{F}_{\rm b,ex}$ to account for the electric interaction between the charged ions
\begin{equation}
\beta\mathcal{F}_b[\{\rho_i\}]=\beta\mathcal{F}_{\rm b,id}[\sigma]+\beta\mathcal{F}_{\rm b,ex}[\sigma]=\sum_i\int_\mathcal{R} \dif^3 {\bf r }\rho_i({\bf r})[\log\left(\rho_i({\bf r})\Lambda_i^3\right)-1]+\frac{1}{2}\int_\mathcal{R} \dif^3 {\bf r}\, \rho_e({\bf r\,})\phi({\bf r}), 
\end{equation}
with $\rho_e({\bf r})=\sum_i z_{i}\rho_i({\bf r})$ the net charge density, $\beta=(k_{\rm B}T)^{-1}$ the inverse thermal energy, $\phi({\bf r},t)=\beta e\psi({\bf r},t)$ the dimensionless electrostatic potential and $e$ the proton charge. 

For the surfaces charges, we can set up a analogous free energy for $\sigma$, $\mathcal{F}_s$, which is also the sum of an ideal part, $\mathcal{F}_{\rm s,id}$, where we must take into account that the surface charges are confined to move on a lattice (multiply occupied sites are forbidden), and an electrostatic excess functional $\mathcal{F}_{\rm s,ex}$,
\begin{equation}
\begin{aligned}
\beta\mathcal{F}_s[\sigma]=&\beta\mathcal{F}_{\rm s,id}[\{\rho_i\}]+\beta\mathcal{F}_{\rm s,ex}[\{\rho_i\}]\\
=&\int_{\mathcal{S}} \dif^2{\bf r}_s \Big{[} {}\sigma({\bf r_s})\log\left(\frac{\sigma({\bf r_s})}{\Gamma}\right)+
\left(\Gamma-\sigma({\bf r}_s)\right)\log\left(\frac{\Gamma-\sigma({\bf r}_s)}{\Gamma}\right)\Big{]}+\frac{1}{2}z_{\sigma}\int_{\mathcal{S}} \dif^2{\bf r}_s \ \sigma({\bf r}_s)\phi({\bf r}_s),
\end{aligned} \label{eq:surffunc}
\end{equation}
where $\Gamma$ is the total density of chargeable sites and $z_{\sigma}$ the valency of the surface charges.  We include no free energy of binding included in Eq. (\ref{eq:surffunc}), since we are interested in out-of-equilibrium processes. We are interested in the chemical desorption reaction SC $\rightleftharpoons$ S$^-$+C$^+$, which we can describe by the rate equation
\begin{equation}
\frac{\dif \{\mathrm{S}\mathrm{C}\}}{\dif t}=k^\text{des}\{\mathrm{S}\mathrm{C}\}-k^\mathrm{ads}[\mathrm{C}^+]\{\mathrm{S}^-\}. \label{eq:rate}
\end{equation}
Here $k^{\rm ads}$ is the adsorption rate constant and $k^{\rm des}$ is the desorption rate constant. We use curly brackets to indicate surface densities, and square brackets to indicate volume densities. In equilibrium, the time derivative vanishes and we obtain the Langmuir adsorption isotherm with chemical reaction constant$K\equiv\{\mathrm{S}^-\}[\mathrm{C}^+]/\{\mathrm{SC}\}=k^\text{des}/k^\text{ads}$

The continuity equation for the ionic species is given by
\begin{equation}
\frac{\dif\rho_i({\bf r},t)}{\dif t}=-\nabla\cdot{\bf J}_i({\bf r\,},t),\quad{\bf r}\in\mathcal{R}\label{eq:contbulk},
\end{equation}
with ${\bf J}_i$ the bulk flux of ion species $i$, where we should note that we used the full (material) derivative of $\rho_i({\bf r},t)$ in order to account for advection. The bulk current ${\bf J}_{i}({\bf r},t)$ can be derived using DDFT,
\begin{equation}
{\bf J}_{i}({\bf r},t)=-D_{b,i}\rho_i({\bf r},t)\nabla\left(\left. \frac{\delta\beta\mathcal{F}_b\left[\rho_{i}\right]}{\delta\rho_{i}({\bf r})}\right\vert_{\substack{\rho_{i}({\bf r},t)}}\right)=-D_{b,i}\left(\nabla\rho_i({\bf r},t)+z_i\rho_i({\bf r},t)\nabla\phi({\bf r},t)\right), 
\label{eq:bulkcurrent}
\end{equation}
where $D_{b,i}$ are diffusion coefficients for the ions in the liquid. The continuity equation for $\sigma({\bf r}_s)$ includes a source/sink term in order to account for the chemical reaction,
\begin{equation}
\frac{\partial\sigma({\bf r}_s,t)}{\partial t}=-\nabla_\mathcal{S}\cdot{\bf j}_{\sigma}({\bf r}_s,t)+ R({\bf r}_s,t),\quad{\bf r}_s\in\mathcal{S}. \label{eq:contsurf}
\end{equation}
where $R$ is the production rate of surface charges and ${\bf j}_{\sigma}$ the (2D) flux of surface charges and $\nabla_{\mathcal{S}}$ is the (2D) divergence in $\mathcal{S}$. For example, for a flat plate in the $xy$-plane, we have $\nabla_\mathcal{S}=(\partial_x,\partial_y)$.  We have implemented a type B dynamic model because the total number of ions (on surface plus in the water) is conserved. From Eq. \eqref{eq:rate} we can write down $R({\bf r}_s)$
\begin{equation}
R({\bf r}_s,t)=-k^\text{ads}\left(\Gamma-\sigma({\bf r}_s,t)\right)+k^\text{des}\rho_{\rm C}({\bf r}_s,t)\sigma({\bf r}_s,t).
\label{eq:rateeq}
\end{equation}
Analogously to the bulk equation, the surface current is given by,
\begin{equation}
\begin{aligned}
{\bf j}_{\sigma}({\bf r}_s,t)&=-D_{s}\sigma({\bf r}_s,t)\nabla_\mathcal{S}\left(\left. \frac{\delta\beta\mathcal{F}_s\left[\sigma\right]}{\delta\sigma({\bf r}_s)}\right\vert_{\substack{\sigma({\bf r}_s,t)}}\right)\\
&=-D_s\left(\dfrac{\Gamma\nabla_\mathcal{S}\sigma({\bf r}_s,t)}{\Gamma-\sigma({\bf r}_s,t)}+z_{\sigma}\sigma({\bf r}_s,t)\nabla_\mathcal{S}\phi({\bf r}_s,t)\right), 
\end{aligned}\label{eq:surfcurrent}
\end{equation}
with $D_s$ the diffusion coefficient for the ions adsorbed in the Stern layer. The final equation needed in order to close the above set is the Poisson equation for the electric potential $\psi$,
\begin{equation}
\nabla^2\phi({\bf r},t)=-4\pi\lambda_B[\rho_e({\bf r},t)+z_{\sigma} \sigma({\bf r},t)f_\mathcal{S}({\bf r})],
\label{eq:poisson}
\end{equation}
where $\lambda_B=\frac{\beta e^2}{4\pi\epsilon}$ is the Bjerrum length with $\epsilon$ the permittivity, and the function $f_{\mathcal{S}}({\bf r})$ (with dimension inverse length) encodes the location of the chargeable surface $\mathcal{S}$. In the case of a chargeable plate parallel to the xy-plane at $z=0$, we have $f_\mathcal{S}({\bf r})=\delta(z)$. Combined with the Navier-Stokes equation for the fluid flow, Eqs. (\ref{eq:contbulk}),(\ref{eq:rateeq}),(\ref{eq:contsurf}),(\ref{eq:bulkcurrent}) \& (\ref{eq:surfcurrent}),(\ref{eq:poisson}) gives the set of governing equations. 

We couple the bulk ions to the surface charges via a Robin boundary condition in $\mathcal{R}$, such that the total influx of counter ions is equal to the destruction rate of the surface charges $-R$. In addition to the standard electric boundary condition and the no-slip boundary condition for ${\bf u}$, we obtain the boundary conditions 
\begin{equation}
\begin{aligned}
-{\bf n}_s\cdot{\bf J}_{\rm C}({\bf r}_s,t)&=-R({\bf r}_s,t),\\
{\bf n}_s\cdot \nabla \phi({\bf r}_s,t)&=-4\pi\lambda_{\rm B} \sigma({\bf r}_s,t)\\
{\bf u}({\bf r}_s,t)&=0
\end{aligned}\label{eq:coupling2}
\end{equation}
with ${\bf n}_s$ an inward pointing normal vector (into the fluid) and ${\bf J_{\rm C}}$ the counter ion flux. All other, non-charged surfaces are impermeable for all ions. The charged surface at $\mathcal{S}$ is impermeable for all ions expect the counter ion.

In order we check the consistency of the governing equations, we consider equilibrium condition where all time derivatives and fluxes vanish. Therefore, we find that in equilibrium $R({\bf r}_s,t)=0$, and Eq. \eqref{eq:rateeq} reduces to the Langmuir adsorption isotherm,
\begin{equation}
\sigma({\bf r_s},t\rightarrow\infty)=\Gamma\left[1+\frac{\rho_C({\bf r_s},t\rightarrow\infty)}{K}\right]^{-1}. \label{eq:langmuir}
\end{equation}
In order for all bulk fluxes to vanish, the functional derivative of $\mathcal{F}$ must reduce to a constant, 
\begin{equation}
\frac{\delta\beta\mathcal{F}_b}{\delta\rho_i({\bf r})}=\text{constant}=\mu_{\mathcal{R}},
\end{equation}
with $\mu_{\mathcal{R}}$ the chemical potential of the system (there is no external potential except for the hard wall potential) fixed by $\rho_{b,i}$, the bulk concentrations of the ions. Solving for the densities $\rho_i$ we get the Boltzmann distribution for all ionic species,
\begin{equation}
\rho_i({\bf r})=A_i\exp[-z_i\phi({\bf r},t\rightarrow\infty)]. \label{eq:boltzmann}
\end{equation}
In the grand canonical ensemble the integration constant $A_i$ is fixed by the chemical potential $\mu_\mathcal{R}$.

Combining Eqs. \eqref{eq:langmuir} and \eqref{eq:boltzmann} we find that
\begin{equation}
\frac{\delta\beta\mathcal{F}_s}{\delta\sigma({\bf r}_s)}=\text{constant},
\end{equation}
and therefore represent a valid equilibrium condition. Alternatively, we could also have enforced ${\bf j}_{\sigma}({\bf r}_s)=0$, and thus a constant first derivative of $\mathcal{F}_s$, which can then be solved for the equilibrium surface charge
\begin{equation}
\sigma({\bf r}_s)=\Gamma\left[1+C_{\sigma} \, e^{z_{\sigma}\phi({\bf r}_s,t\rightarrow\infty)}\right]^{-1}, \quad{\bf r}_s\in\mathcal{S},
\end{equation}
with $C_{\sigma}$ an integration constant. We can then combine this with the condition $R=0$ to obtain once again the Langmuir adsorption isotherm for the surface charge and the Boltzmann weight for the dissolved ions. This shows the internal consistency of the theory, for if we set any 2 of ${\bf J}_i$, ${\bf j}_{\sigma}$ or $R$ to zero, it follows that the third vanishes.

\section{Electric Body Force}\label{sec:EBF}

\begin{figure}[!ht]
\centering
\includegraphics[width=0.6\textwidth]{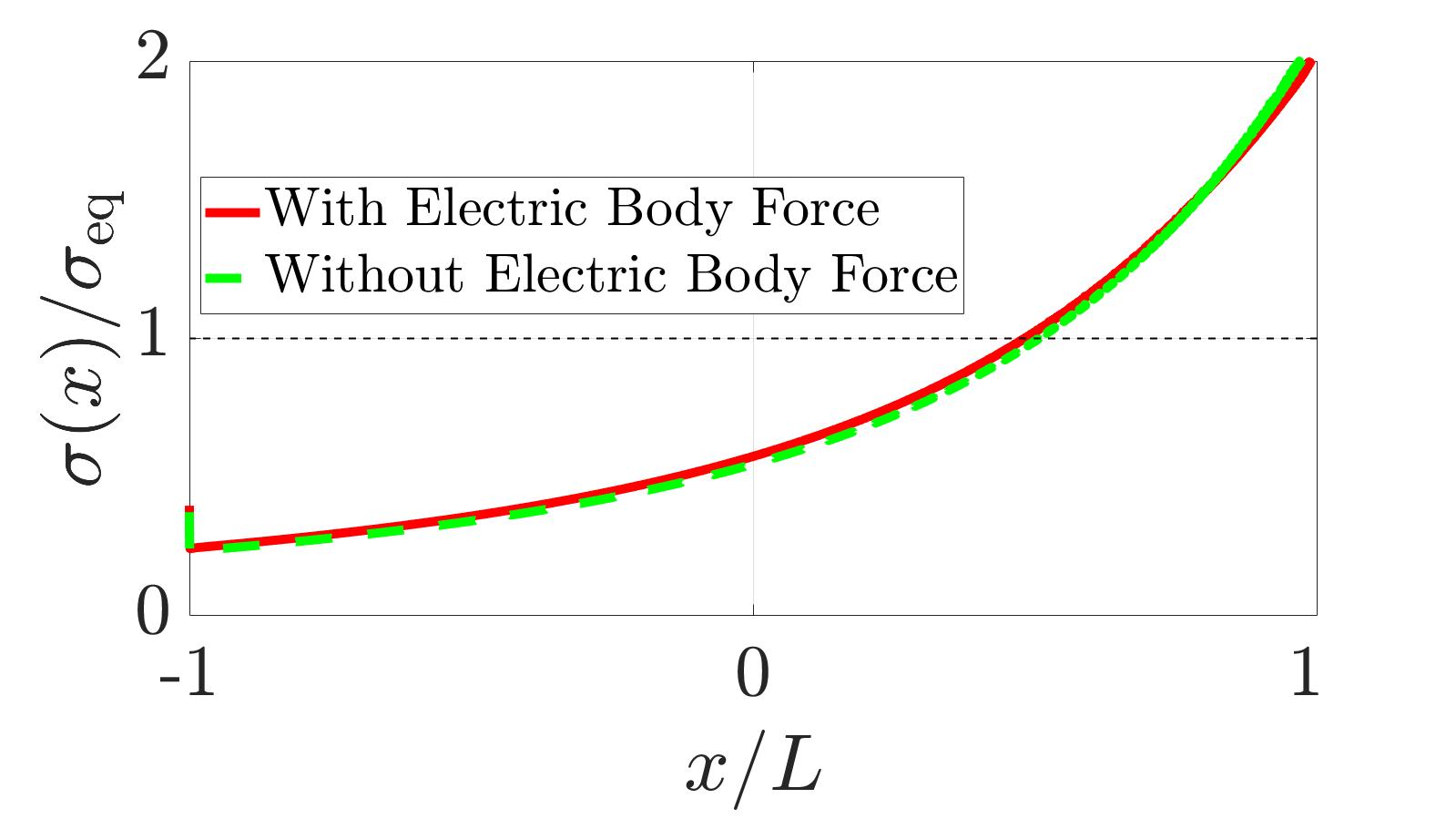}
\caption{The surface charge profile at steady state in the conduction-dominated regime with (red full line) and without (green dashed line) the electric body force included in the Navier-Stokes equation. The channel length $2L=30 \,\mu$m, channel height $H=1\,\mu$m and a pressure drop $\Delta p=0.5\,$bar was used.}\label{fig:EBF}
\end{figure}

In section \ref{sec:PNP} we assumed that the electric body force in the Navier-Stokes equation is negligible in pressure-driven electrokinetic systems. The electric body force is proportional to the local net charge density, and therefore is only non-zero in the Electric Double Layer. Although comparable in magnitude, the pressure gradient extends through the whole channel, and since we have that $H\gg\lambda_D$ we can therefore a priori expect that the electric body force will have a negligible effect compared to the applied pressure gradient. To test this assumption, we numerically calculated the surface charge profile in the conduction-dominated regime, shown in Fig. \ref{fig:EBF}. 
In the conduction-dominated regime, the electric field is large and thus the electric body force is large. Regardless, Fig. \ref{fig:EBF} shows that including the electric body force has no significant impact on the surface charge profile at steady state. We can thus safely neglect the electric body force and still obtain the same qualitative behaviour.

\section{Derivation of zeta potential and net charge current}\label{sec:calculations}

The integration constant $\zeta(x)$ of the solution to the charge density $\rho_e(x,z)$, Eq. (\ref{eq:rhoe}), can be determined using Eq. (\ref{eq:coupling}). We rewrite the counter ion flux as $\hat{z}\cdot{\bf J}_C(x,0)=z_C\hat{z}\cdot{\bf J}_e(x,0)$ and use the expression for the net charge flux ${\bf J}_e$, Eq. (\ref{eq:Je}), and solution to $\rho_e$, Eq. (\ref{eq:rhoe}), to find
\begin{equation}
z_C\hat{z}\cdot{\bf J}_e(x,0)=z_{\sigma}D\left(\lambda_D^{-1}\rho\zeta(x)+4\pi\lambda_B\rho\sigma(x)\right)=k^{\rm des}\Gamma-\left(k^{\rm des}+k^{\rm ads}\rho_{C,s}\right)\sigma(x),
\end{equation}
where we have used the electrostatic boundary condition ${\bf n}_s\cdot\nabla \psi_s=-\sigma/\epsilon$ to eliminate the electrostatic potential $\psi$. Now we can use the solution to $\sigma$, Eq. (\ref{eq:sigma}), to derive an expression for $\zeta(x)$,
\begin{equation}
\begin{aligned}
z_{\sigma}\frac{D}{\lambda_D}\rho\zeta(x)&=k^{\rm des}\Gamma-\left(k^{\rm des}+k^{\rm ads}\rho_{C,s}-D\lambda_D^{-2}\right)\sigma(x),\\
&=z_{\sigma}\frac{D}{\lambda_D}\rho\zeta_{\rm eq}+\sigma_{\rm eq}(\tau_{\rm EDL}^{-1}-\tau_{\rm reac}^{-1})(a_+ e^{k_+x}+a_-e^{k_-x})
\end{aligned}
\end{equation}
Here, we used that $k^{\rm des}\Gamma-\left(k^{\rm des}+k^{\rm ads}\rho_{C,s}\right)\sigma_{\rm eq}=0$ by definition and identified the expression for the dimensionless equilibrium zeta potential $\zeta_{\rm eq}=z_{\sigma}4\pi\lambda_B\lambda_D\sigma_{\rm eq}$ from linear Poisson-Boltzmann theory. As argued in the text, we know that $\tau_{\rm EDL}=\lambda_D^2/D \gg\tau_{\rm reac}$. Using the identity $\rho\lambda_D=(4\pi\lambda_B\lambda_D)^{-1}$ we can rewrite the expression for $\zeta(x)$ as
\begin{equation}\label{eq:zetaappendix}
\zeta(x)=\zeta_{\rm eq}+\frac{\sigma_{\rm eq}}{z_{\sigma}\rho\lambda_D}\left(a_+ e^{k_+x}+a_-e^{k_-x}\right)=\zeta_{\rm eq}\left(1+a_+ e^{k_+x}+a_-e^{k_-x}\right).
\end{equation}

To determine $E$, we impose that at any position $x$ no net current passes through any channel slice with normal $\bf \hat{x}$. This condition is a direct consequence of the vanishing divergence of ${\bf J}_e$ and the open circuit geometry. Using the symmetry of the system we thus demand that
\begin{equation}\label{eq:nonetcurrent}
z_{\sigma}j_{\sigma}(x)+\int_0^{\frac{1}{2}H} \dif z J_{e,x}(x,z)=0,
\end{equation}
where $z_{\sigma}j_{\sigma}(x)$ is the net charge current through the Stern layer. Using Eq. (\ref{eq:Je}), and the solution to $\rho_e$, Eq. (\ref{eq:rhoe}), we find the net charge current through the liquid
\begin{equation}\label{eq:netje}
\begin{aligned}
\int_0^{\frac{1}{2}H} \dif z J_{e,x}(x,z)&=-D\partial_x\int_0^{\frac{1}{2}H} \dif z \rho_e+ z_{\sigma}D\rho\int_0^{\frac{1}{2}H} \dif z E+\frac{\partial_x p}{2\eta}\int_0^{\frac{1}{2}H} \dif z z(H-z) \rho_e,\\
&\approx 2D\rho\left(\lambda_D\partial_x\zeta(x)+\frac{1}{2}z_{\sigma}EH-\frac{\partial_xpH\lambda^2_D}{2D\eta}\zeta(x)\right)
\end{aligned}
\end{equation}
where have used that $H\gg \lambda_D$ and included a factor $z_{\sigma}$ in the second term on the right hand side since $E$ was defined as a positive quantity. Lastly, we use Eq. (\ref{eq:sigma}) and (\ref{eq:bc2}) to derive an expression for the Stern-layer current 
\begin{equation}\label{eq:jsig}
j_{\sigma}(x)=-D_s(\partial_x\sigma-E\sigma)=D_sE\sigma_{\rm eq}+D_sk_-a_+e^{k_+x}+D_sk_+a_-e^{k_-x},
\end{equation}
where we used that $E-k_{\pm}=k_{\mp}$. Eqs. (\ref{eq:zetaappendix}), (\ref{eq:netje}) and (\ref{eq:jsig}) can then be plugged in Eq. (\ref{eq:nonetcurrent}) which can then in principle be solved for the streaming electric field $E$. 

\end{document}